**Emerging-properties Mapping Using Spatial Embedding Statistics: EMUSES**

Chris Foulon[1,2], Marcela Ovando-Tellez[1,2], Lia Talozzi[2,3], Maurizio Corbetta[4,5,6], Anna Matsulevits[1,2], Michel Thiebaut de Schotten[1,2]

[1]Groupe d'Imagerie Neurofonctionnelle (GIN), Institut des Maladies Neurodegeneratives-UMR 5293, CNRS, Bordeaux, France,
[2]Brain Connectivity and Behaviour Laboratory, Sorbonne Universities, Paris, France
[3]Stanford Medical School, Stanford, CA
[4]Clinica Neurologica, Department of Neuroscience, University of Padova, Padova, 32122, Italy
[5]Padova Neuroscience Center (PNC), University of Padova, Padova, 32122, Italy
[6]Venetian Institute of Molecular Medicine, VIMM, Padova, 32122, Italy

# Abstract

Understanding complex phenomena often requires analyzing high-dimensional data to uncover emergent properties that arise from multifactorial interactions. Here, we present EMUSES (Emerging-properties Mapping Using Spatial Embedding Statistics), an innovative approach employing Uniform Manifold Approximation and Projection (UMAP) to create high-dimensional embeddings that reveal latent structures within data. EMUSES facilitates the exploration and prediction of emergent properties by statistically analyzing these latent spaces. Using three distinct datasets—a handwritten digits dataset from the National Institute of Standards and Technology (NIST, E. Alpaydin, 1998), the Chicago Face Database (Ma et al., 2015), and brain disconnection data post-stroke (Talozzi et al., 2023)—we demonstrate EMUSES' effectiveness in detecting and interpreting emergent properties. Our method not only predicts outcomes with high accuracy but also provides clear visualizations and statistical insights into the underlying interactions within the data. By bridging the gap between predictive accuracy and interpretability, EMUSES offers researchers a powerful tool to understand the multifactorial origins of complex phenomena.

# Introduction

Empirical observations have long demonstrated that the whole is often greater than the sum of its parts, a concept first articulated by Aristotle in 350 BCE (Hamlyn, 1984). With the advent of large open datasets, scientists have confirmed that pathologies frequently arise from the interplay of multiple environmental, genetic, and biological factors (e.g., Balestri et al., 2024; Virolainen et al., 2023). This principle is also evident within individual modalities, where, for instance, cognition and pathology are thought to emerge from the integration or disruption of interactions between brain regions (Pacella et al., 2019; Pessoa, 2023;

Thiebaut de Schotten & Forkel, 2022). Advanced genetic research further illustrates that a combination of genetic factors determines characteristics ranging from normal traits, such as grain size in rice (Ngangkham et al., 2018), to complex pathologies like schizophrenia (Jiang et al., 2023). While the multifactorial origins of function and disorders are intuitively recognized, precise methods for statistically identifying these interactions in rich datasets are still missing (Pearl, 2019).

Numerous advanced statistical and computational methods are available to quantify complex interactions in rich datasets. These include spatial embedding techniques that cluster covariate factors along primary explanatory dimensions (i.e., a latent space), such as Principal Component Analysis (PCA; Jolliffe, 2002), Independent Component Analysis (ICA; Hyvärinen & Oja, 2000), t-Distributed Stochastic Neighbor Embedding (t-SNE; Maaten & Hinton, 2008), Uniform Manifold Approximation and Projection (UMAP; McInnes et al., 2020), and autoencoders (Goodfellow, 2016). Despite the availability of these methods, integrating and interpreting data from multiple factors remains a statistically significant hurdle. Inherent challenges persist in associating the complexity of these multifactorial interactions with function or disorder. Machine learning and artificial intelligence are increasingly employed to circumvent these limitations and uncover patterns and interactions that traditional statistical methods might miss. However, while these approaches show promising results in terms of predictive power, they are still limited regarding explanatory power. New methods are now required to explore the mechanisms underpinning the emergence of complex cognitive functions, which seem to evade localized or sequential explanations (Thiebaut de Schotten & Forkel, 2022).

Additionally, it is important to consider that "the same thing can come to be in more than one way," (Hamlyn, 1984). Most statistical investigations, however, tend to favor identifying a single cause associated with or predicting an observation (Pearl, 2019). For example, memory disorders can manifest due to different non-overlapping lesions localized along interconnected regions within the limbic system (Alves et al., 2019; Catani et al., 2013). Therefore, it's necessary to employ more sophisticated approaches to address the complexity of multiple causes leading to the same outcomes (Pearl, 2019).

In this work, we introduce a novel approach, EMUSES (Emerging-properties Mapping Using with Spatial Embedding Statistics), which leverages UMAP's capability to capture complex interactions within data. EMUSES statistically explores the latent space to identify patterns associated with variables of interest and predict outcomes for new out-of-sample data. EMUSES is a free, open-source, and user-friendly python tool for researchers and practitioners to investigate the emergent properties of complex datasets. We demonstrate the performance and interpretability of EMUSES using three diverse datasets. First, with the NIST (National Institute of Standards and Technology) dataset of 1797 handwritten digits (E. Alpaydin, 1998), we showcase EMUSES' reliability in capturing and predicting intricate patterns. Next, we apply EMUSES to the Chicago Face Database, containing 685 pictures of faces (Ma et al., 2015), to explore its ability to capture and predict subjective measures such as attractiveness. Finally, we validate EMUSES on brain disconnection data, comparing its performance with the existing Disconnectome Symptom Discoverer (Talozzi et al., 2021), which inspired EMUSES' development.

## Results

First, embedding the handwritten digits (NIST, E. Alpaydin, 1998) displayed a latent space in a highly clustered structure (**Figure 1a**), revealing redundancy in how digits can be written. We explored whether this space can capture the conceptual meaning of handwritten digits, defined through the interaction of pixels forming shapes that emerge as numbers. For example, EMUSES was able to define two distinct territories—one additional territory was identified but did not have a sufficient effect size—for the digit "9", demonstrating that the concept of the number 9 can emerge from two different handwritten configurations (**Figure 1b**). EMUSES performed a Mann-Whitney comparison between images inside and outside the clusters to understand these configurations better. Results indicated that the typical handwritten shape of the number nine could be identified as either the European or more East-Asian style (**Figure 1b;** 'Regional Handwriting Variation', 2024). Finally, new handwritten digits, previously unseen, were projected into the latent space and subsequently identified by the random forest trained within EMUSES with a 93.33% class prediction accuracy (Confusion matrix in **Figure 1c**).

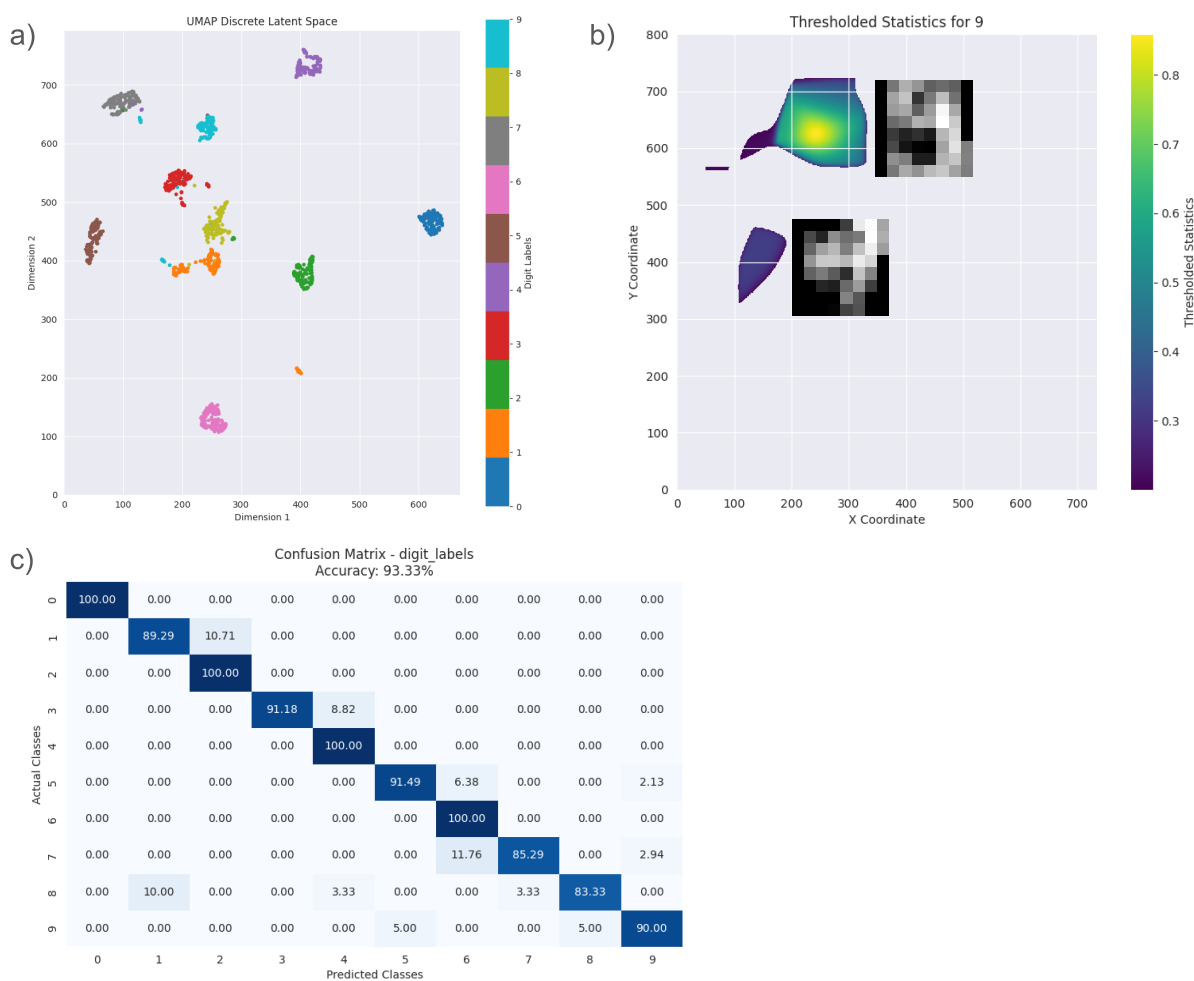

**Figure 1.** EMUSES applied to handwritten digits (**a**) latent space plot showing a highly clustered structure of handwritten digits (color corresponds to the label) (**b**) Two distinct clusters with medium to large effect sizes for the digit '9', indicating European and East-Asian handwriting styles. (**c**) Prediction of unseen handwritten digits embedded into the latent space and classified with Random Forest.

Next, the embedding of facial features from the Chicago Face Database (Ma et al., 2015) revealed two heavily clustered structures (**Figure 2a**). We investigated whether the latent space could capture attractiveness as evaluated by independent raters. attractiveness, being an emergent property influenced by many factors, was effectively captured by EMUSES. The tool revealed that multiple sets of facial features contribute to attractiveness. A higher attractiveness cluster was primarily explained by the combination of the asymmetry of the pupils distance to the top of the head (z-scored U = 7.23, P < 0.001, effect size = 0.31), how heart-shaped the face is (z-scored U = 6.78, P < 0.001, effect size = 0.29), the upper head length (z-scored U = 5.27, P < 0.001, effect size = 0.23) and the forehead height (z-scored U

= 7.48, P < 0.001, effect size = 0.32). A second cluster with lower attractiveness driven mostly by the global shape of the face with the face shape or cheek prominence compared with face length (z-scored U = 5.70, P < 0.001, effect size = 0.24), face roundness (z-scored U = 5.48, P < 0.001, effect size = 0.23) and width to height ratio (z-scored U = 5.69, P < 0.001, effect size = 0.24) (**Figure 2b**). However, a random forest trained on the embedding coordinates to predict attractiveness only showed a 0.09% correlation of the prediction over real scores on unseen face characteristics (**Figure 2c**).

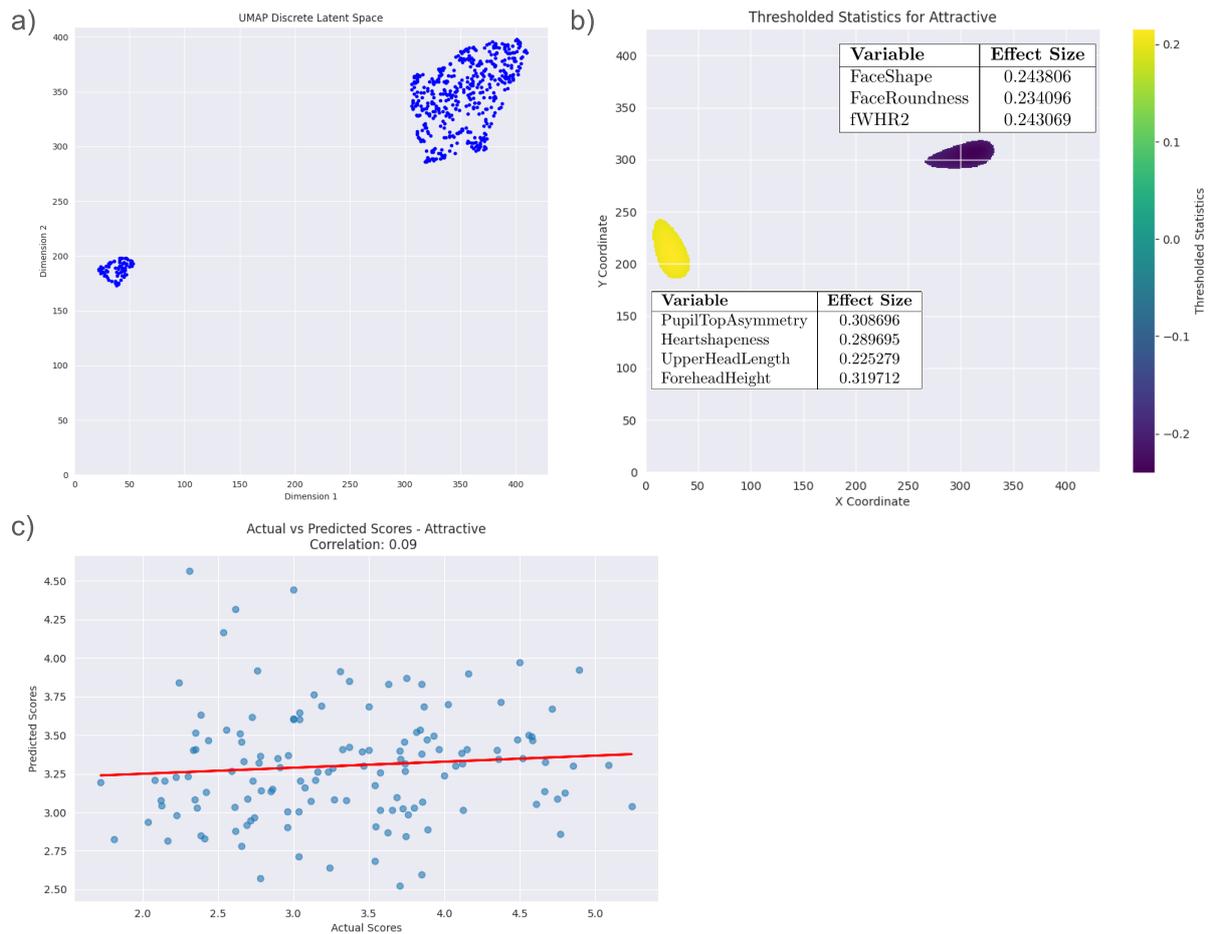

**Figure 2.** EMUSES applied to faces (**a**) latent space plot of facial features from the Chicago Face Database, showing clustered structures. (**b**) Two distinct clusters representing different sets of facial features, the left one contributing positively and the right one positively to attractiveness. (**c**) Random Forest attractiveness prediction performance on out of distribution embedded facial characteristics.

Finally, we replicated part of the analyses by Talozzi et al. (2021) to produce an embedding of brain disconnection patterns after stroke, using 1333 stroke disconnection maps (i.e., disconnectomes; Foulon et al., 2018) . This replication produced a latent space similar to that described in Talozzi et al. (2021), where disconnection patterns displayed a partially segregated structure (**Figure 3a**). Similar to the conceptual emergence seen in handwritten digits and facial attractiveness, the neuropsychological profiles of patients emerged from the combination of different networks interrupted by their lesions. Unlike the other datasets, the 1333 stroke disconnection maps did not include neuropsychological assessments. Instead, we projected a second dataset of 119 extensively assessed cognitive profiles into the latent space to enrich it with neuropsychological data. This approach allowed us to leverage the larger dataset to define the latent space based on disconnection profiles and use the smaller dataset to explore neuropsychological associations within that space. For example, EMUSES identified two distinct territories associated with Mesulam centre of cancellation (CoC; M-M, 1985), demonstrating that visual neglect could emerge from disconnection patterns in either the left or right hemisphere, leading to similar deficits in this neuropsychological test (**Figure 3b**). Subsequently, 20 new patients' disconnectomes were projected into the latent space, predicting their neuropsychological performance (i.e., CoC) with 65% correlation between the predicted and actual scores (**Figure 3c**).

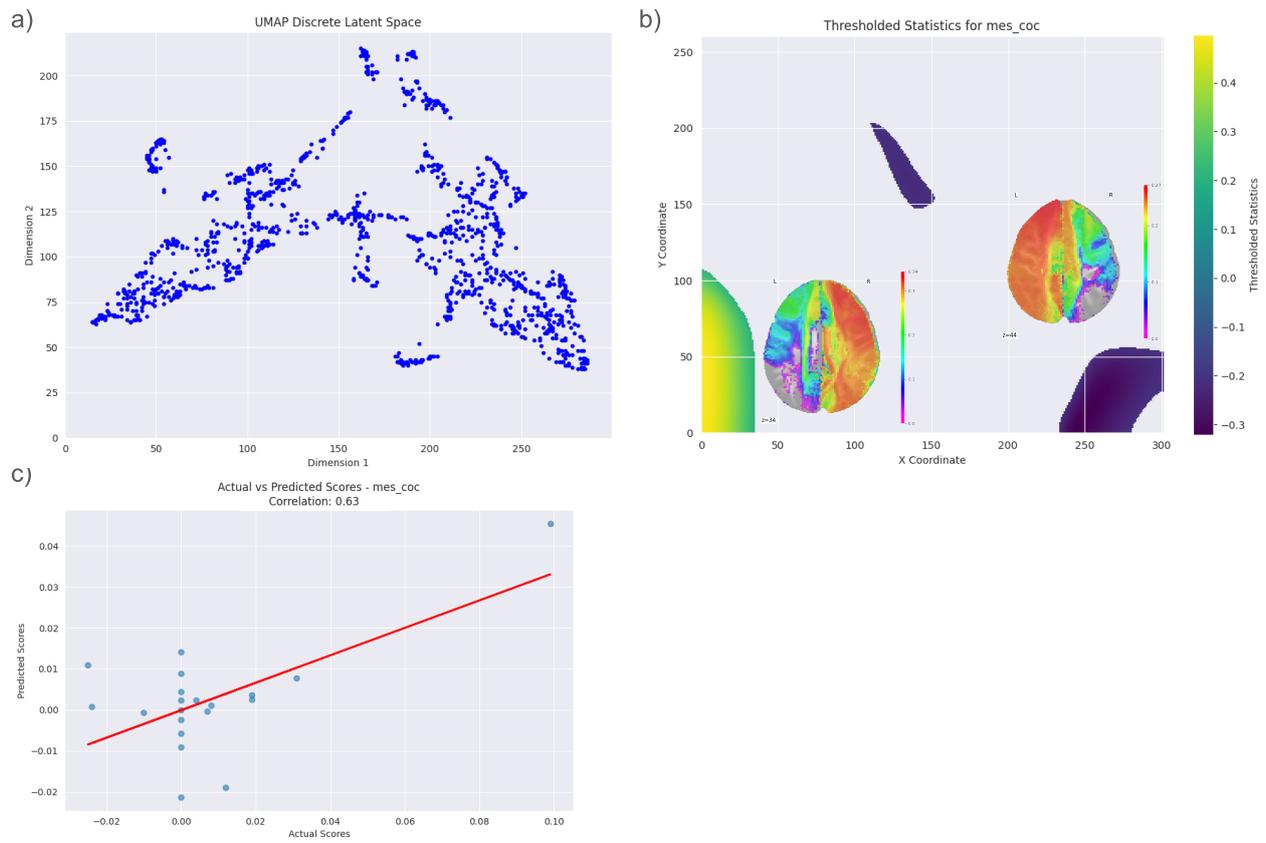

**Figure 3.** EMUSES applied to clinical neuroimaging (**a**) latent space plot of brain disconnection patterns after stroke, showing a partially segregated structure. (**b**) Two distinct clusters associated with visual neglect, indicating different patterns of disconnection in the left or right hemisphere leading to similar deficits. The middle cluster's did not have a sufficient effect size to extract a pattern (**c**) Performance of Random Forest CoC prediction from unseen disconnectome maps.

## Discussion

We presented EMUSES (Emerging-properties Mapping Using with Spatial Embedding Statistics), an innovative approach that leverages high-dimensional embedding to statistically capture intricate interactions within data. We verified its effectiveness using three distinct datasets. Our findings show that EMUSES detects and forecasts emergent properties in complex datasets. Additionally, it provides a fresh statistical perspective on the exploration of the multifaceted nature of various phenomena.

In the three examples provided, EMUSES effectively detected, identified, and predicted combinations of features (i.e., pixels' arrangement, facial features, or brain disconnections) leading to specific observations, such as the concept of a number, attractiveness, or neuropsychological deficit. These results illustrate that our method is well-suited for exploring emerging properties in various datasets. Although predicting outcomes based on feature combinations is common in machine learning and artificial intelligence (Vieira et al., 2022), our approach offers the distinct advantage of providing visualization and statistical detection, which enhances data interpretability. Visualizing latent spaces and statistically outlining significant combinations of features allows for a better comprehension of the underlying structures and interactions within the data, thereby enabling more accurate interpretations. By capturing the intricate interplay between various factors, EMUSES bridges the gap between predictive accuracy and explanatory power, a limitation often encountered with traditional machine learning methods.

EMUSES also provides a methodological advancement in identifying and understanding the complexity arising from the multifactorial origins of many phenomena. As illustrated in the three examples provided, EMUSES successfully identified and explained each observation based on multiple causal factors. By effectively uncovering and identifying multiple explanations, EMUSES provides a robust tool for exploring the nuanced mechanisms underlying various observations, thereby advancing our understanding of complex systems

As a note of limitation, the results showcased in this manuscript were systematically based on a 2-dimensional embedding of features for practical visual reasons. Additionally, default embedding parameters were used throughout the analysis. However, EMUSES is designed to support higher-dimensional embeddings, and optimizing parameters specific to each dataset could offer a more accurate representation of the data distribution. Exploring these higher-dimensional embeddings and parameter optimizations may reveal more complex interactions within the data, enhancing the depth and accuracy of the analysis.

Future work should broaden the usage of EMUSES to encompass diverse and intricate datasets, including but not limited to genomic data, social networks, and ecological systems. Furthermore, integrating EMUSES with advanced statistical and machine learning techniques will strengthen its capacity to unravel complex interactions. The continuous advancement and validation of EMUSES in real-world scenarios are crucial to solidifying its effectiveness and dependability across scientific and practical applications.

# Methods

## 2.1 Datasets

We selected three datasets of varied nature to demonstrate the flexibility and capabilities of the EMUSE. First, we used the NIST handwritten digit dataset containing 1797 grayscale size-normalized and centered images of 8x8 pixels with associated digit labels (0 to 9). For the second analysis, we used 43 objective numerical facial features of the Chicago Face Database (e.g. nose length, eye width, chin length, and so on) to explore subjective attractiveness—from 1087 raters on a 1-7 Likert scale where 1 means "Not at all attractive" and 7 means "Extremely attractive". We only used the objective features with data for the whole dataset and removed the "CheekboneProminence" feature, which contained inconsistent values. Finally, we explored the impact of human brain disconnections on behavior using the datasets described in Talozzi and colleagues' study (Talozzi et al., 2021). We used the 1333 disconnectome maps (Foulon et al., 2018)—estimation maps of the probability of white matter tract disconnections in the healthy population based on the patient's lesion—to create the UMAP latent space and then 119 and 20 disconnectomes with their corresponding scores for correlations and predictions. We investigated the disconnection patterns associated with the 87 scores, encompassing a wide range of cognitive domains.

## 2.2 Preprocessing and UMAP training

To be entered in UMAP, we need to create a matrix with the observations as rows and the features as columns. As we used a dataset of 2D images and another with 3D images, we flattened these data into vectors where the value of a pixel or a voxel corresponds to a feature. UMAP is sensitive to the scale of the features; therefore, we need to ensure all the features are on the same scale. While the image datasets do not require scaling, the facial attributes have various scales—e.g. lengths in millimeters that are not comparable between features—so we rescaled the values between 0 and 1. We then trained a UMAP, generating a latent space for each dataset. With the digits, we created the latent space using the 1437 digits images (training split) and used the 360 remaining (test split) for the statistics and prediction. We also used an 80/20% split (548/137) for the facial features dataset to explore the subjective attractiveness. Finally, we used the 1333 disconnectome maps (which did not have psychological scores) to create the embedding space, 119 disconnectomes with scores to perform the statistics and 20 to evaluate the prediction accuracy.

## 2.3 Discrete Latent Space

To perform correlations and statistics, we need a way to quantify the notion of proximity and how it relates to a variable of interest. However, the UMAP embedding space is continuous and unbounded. We implemented an automated discretization of the latent space to alleviate this issue. First, we rescale the embedding space between 0 and 1 and then choose a pixel resolution. Instead of empirically selecting a resolution, which would require figuring out an aspect ratio, we algorithmically compute the best resolution for a given percentage of overlap between the embedding points. Indeed, when discretizing the space, we need to round the continuous embedding coordinates, which leads to overlapping points. Although we could select a space without overlap, the pixel resolution would prevent any analysis due to unreasonable computing demand. We end up with a Discrete Latent Space (DLS) where each embedding coordinate has a corresponding pixel—here, we chose a 2D latent space, but EMUSES can create and use N-Dimensions spaces—with a value of 1, the rest of the pixels being 0.

## 2.4 Extract spatial patterns related to the variable of interest

As the goal of the statistics is to relate the embedding spatial patterns with a variable of interest, we create this DLS with the part of the dataset with a value for this variable. In our case, the digits and faces datasets have values for the whole dataset. However, the disconnectome dataset used to create the latent space does not have psychological assessments. We thus used the second dataset, embedded with the already trained UMAP, to create a DLS. With the discretized space, we can now represent proximity using a Gaussian smoothing with the adjustable decay rate—$\sigma$, the standard deviation of the Gaussian distribution determines the rate of decay and the extent of the proximity ($4\sigma$)—depending on the size of the DLS. To allow for straightforward statistical analyses, we add a third dimension to the DLS so each slice contains information about proximity with one embedded coordinate at a time. Then, for each coordinate in the 2D DLS, we have a vector of proximity values with each embedded point. Finally, we can use Pearson (or Point Biserial; Kornbrot, 2014) correlations to evaluate how the vectors of proximity relate to the variable of interest—using a minimum r coefficient of 0.2 to extract the medium to large effects. We used the Point Biserial correlations, which are more adapted to compare continuous with binary variables, for the numbers as the variable of interest is categorical, considering each number as a separate variable of interest for which one means the data point is the digit and zero for another digit.

## 2.5 Exhibit the properties emerging from the data

Correlations on the 3D DLS allow us to extract the often multiple patterns (i.e. clusters) in the embeddings that relate positively or negatively to the variable of interest. We can now extract the prototypical characteristics present in each cluster to easily visualize the different profiles in the data associated with the variable of interest. To come back to the original data, we extract all the coordinates within each cluster and perform a Mann-Whitney non-parametric test from which we compute the effect size—computing an r coefficient from the standardized z-scored U-value (Tomczak & Tomczak, 2014)—for each feature of all observations within the cluster versus those outside. The effect size maps have the same shape as the input data and are the stereotypical profiles of features in each cluster.

## 2.6 Scores prediction from embedding coordinates

EMUSES also provides a way to predict the scores of new data points never seen by the model. We used the versatile and scalable Random Forest (Breiman, 2001) algorithm for this proof of concept. For each variable of interest, we train a Random Forest using 100 permutations of the training embeddings of a 5-fold cross-validation scheme to ensure robust model selection. We then use the permutation with the best global performance on the validation sets on the test set. We ensemble the prediction of the five models trained during the 5-fold cross-validation, producing an average prediction on the out-of-distribution test set. The test set contains only unseen data points embedded in the already trained UMAP space to evaluate the quality and generalization of the prediction. The models trained using 5-fold cross-validation can be loaded into EMUSES to make new predictions.

## 2.7 External code acknowledgement and resources availability

To implement the tool and the analyses, we used the following Python packages: bcblib (https://github.com/chrisfoulon/BCBlib), scikit-learn (Pedregosa et al., 2011), numpy (Harris et al., 2020), scipy (Virtanen et al., 2020) and pandas (team, 2024). We also used matplotlib (Hunter, 2007) and seaborn (Waskom, 2021) for visualization. The code, analyses and future updates will be available on GitHub (https://github.com/chrisfoulon).

**Acknowledgement**

This work is supported by HORIZON- INFRA-2022 SERV (Grant No. 101147319) "EBRAINS 2.0: A Research Infrastructure to Advance Neuroscience and Brain Health", by the European Union's Horizon 2020 research and innovation programme under the European Research Council (ERC) Consolidator grant agreement No. 818521 (DISCONNECTOME), the University of Bordeaux's IdEx "Investments for the Future" program RRI "IMPACT", and the IHU "Precision & Global Vascular Brain Health Institute – VBHI" funded by the France 2030 initiative (ANR-23-IAHU-0001).

**References**

Aristotle. (1984). *The complete works of Aristotle: The revised Oxford translation* (J. Barnes, Ed.). Princeton University Press. (Original work published ca. 350 B.C.E.)

Alves, P. N., Foulon, C., Karolis, V., Bzdok, D., Margulies, D. S., Volle, E., & Thiebaut de Schotten, M. (2019). An improved neuroanatomical model of the default-mode network reconciles previous neuroimaging and neuropathological findings. *Communications Biology*, *2*(1), Article 1. https://doi.org/10.1038/s42003-019-0611-3

Balestri, W., Sharma, R., da Silva, V. A., Bobotis, B. C., Curle, A. J., Kothakota, V., Kalantarnia, F., Hangad, M. V., Hoorfar, M., Jones, J. L., Tremblay, M.-È., El-Jawhari, J. J., Willerth, S. M., & Reinwald, Y. (2024). Modeling the neuroimmune system in Alzheimer's and Parkinson's diseases. *Journal of Neuroinflammation*, *21*(1), 32. https://doi.org/10.1186/s12974-024-03024-8

Breiman, L. (2001). Random Forests. *Machine Learning*, *45*(1), 5–32. https://doi.org/10.1023/A:1010933404324

Catani, M., Dell'Acqua, F., & Thiebaut de Schotten, M. (2013). A revised limbic system model for memory, emotion and behaviour. *Neuroscience & Biobehavioral Reviews*, *37*(8), 1724–1737. https://doi.org/10.1016/j.neubiorev.2013.07.001

E. Alpaydin, C. K. (1998). *Optical Recognition of Handwritten Digits* [dataset]. UCI Machine Learning Repository. https://doi.org/10.24432/C50P49

Foulon, C., Cerliani, L., Kinkingnéhun, S., Levy, R., Rosso, C., Urbanski, M., Volle, E., & Thiebaut de Schotten, M. (2018). Advanced lesion symptom mapping analyses and implementation as BCBtoolkit. *GigaScience*, *7*(3), giy004. https://doi.org/10.1093/gigascience/giy004

Goodfellow, I. (2016). *Deep learning*. Cambridge, Massachusetts : The MIT Press. http://archive.org/details/deeplearning0000good

Hamlyn, D. W. (1984). *Metaphysics*. Cambridge University Press.

Harris, C. R., Millman, K. J., van der Walt, S. J., Gommers, R., Virtanen, P., Cournapeau, D., Wieser, E., Taylor, J., Berg, S., Smith, N. J., Kern, R., Picus, M., Hoyer, S., van Kerkwijk, M. H., Brett, M., Haldane, A., del Río, J. F., Wiebe, M., Peterson, P., … Oliphant, T. E. (2020). Array programming with NumPy. *Nature*, *585*(7825), 357–362. https://doi.org/10.1038/s41586-020-2649-2

Hunter, J. D. (2007). Matplotlib: A 2D Graphics Environment. *Computing in Science & Engineering*, *9*(3), 90–95. https://doi.org/10.1109/MCSE.2007.55

Hyvärinen, A., & Oja, E. (2000). Independent component analysis: Algorithms and applications. *Neural Networks*, *13*(4), 411–430. https://doi.org/10.1016/S0893-6080(00)00026-5

Jiang, Y., Wang, J., Zhou, E., Palaniyappan, L., Luo, C., Ji, G., Yang, J., Wang, Y., Zhang, Y., Huang, C.-C., Tsai, S.-J., Chang, X., Xie, C., Zhang, W., Lv, J., Chen, D., Shen, C., Wu, X., Zhang, B., … Feng, J. (2023). Neuroimaging biomarkers define neurophysiological subtypes with distinct trajectories in schizophrenia. *Nature Mental Health*, *1*(3), 186–199. https://doi.org/10.1038/s44220-023-00024-0

Jolliffe, I. T. (Ed.). (2002). Graphical Representation of Data Using Principal Components. In *Principal Component Analysis* (pp. 78–110). Springer. https://doi.org/10.1007/0-387-22440-8_5

Kornbrot, D. (2014). Point Biserial Correlation. In *Wiley StatsRef: Statistics Reference Online*. John Wiley & Sons, Ltd. https://doi.org/10.1002/9781118445112.stat06227

Ma, D. S., Correll, J., & Wittenbrink, B. (2015). The Chicago face database: A free stimulus set of faces and norming data. *Behavior Research Methods*, *47*(4), 1122–1135. https://doi.org/10.3758/s13428-014-0532-5

Maaten, L. van der, & Hinton, G. (2008). Visualizing Data using t-SNE. *Journal of Machine Learning Research*, *9*(86), 2579–2605.

McInnes, L., Healy, J., & Melville, J. (2020). *UMAP: Uniform Manifold Approximation and Projection for Dimension Reduction* (arXiv:1802.03426). arXiv. https://doi.org/10.48550/arXiv.1802.03426

M-M, M. (1985). Patterns in behavioral neuroanatomy: Association areas, the limbic system, and hemispheric specialization. *Principles of Behavioral Neurology*. https://cir.nii.ac.jp/crid/1571135650193016320

Ngangkham, U., Samantaray, S., Yadav, M. K., Kumar, A., Chidambaranathan, P., & Katara, J. L. (2018). Effect of multiple allelic combinations of genes on regulating grain size in rice. *PLOS ONE*, *13*(1), e0190684. https://doi.org/10.1371/journal.pone.0190684

Pacella, V., Foulon, C., Jenkinson, P. M., Scandola, M., Bertagnoli, S., Avesani, R., Fotopoulou, A., Moro, V., & Thiebaut de Schotten, M. (2019). Anosognosia for hemiplegia as a tripartite disconnection syndrome. *eLife*, *8*, e46075. https://doi.org/10.7554/eLife.46075

Pearl, J. (2019). The seven tools of causal inference, with reflections on machine learning. *Communications of the ACM*, *62*(3), 54–60. https://doi.org/10.1145/3241036

Pedregosa, F., Varoquaux, G., Gramfort, A., Michel, V., Thirion, B., Grisel, O., Blondel, M., Prettenhofer, P., Weiss, R., Dubourg, V., Vanderplas, J., Passos, A., Cournapeau, D., Brucher, M., Perrot, M., & Duchesnay, É. (2011). Scikit-learn: Machine Learning in Python. *Journal of Machine Learning Research*, *12*(85), 2825–2830.

Pessoa, L. (2023). How many brain regions are needed to elucidate the neural bases of fear and anxiety? *Neuroscience & Biobehavioral Reviews*, *146*, 105039. https://doi.org/10.1016/j.neubiorev.2023.105039

Regional handwriting variation. (2024). In *Wikipedia*. https://en.wikipedia.org/w/index.php?title=Regional_handwriting_variation&oldid=1217028705

Talozzi, L., Forkel, S. J., Pacella, V., Nozais, V., Allart, E., Piscicelli, C., Pérennou, D., Tranel, D., Boes, A., Corbetta, M., Nachev, P., & Thiebaut de Schotten, M. (2023). Latent disconnectome prediction of long-term cognitive-behavioural symptoms in stroke. *Brain*, *146*(5), 1963–1978. https://doi.org/10.1093/brain/awad013

team, T. pandas development. (2024). *pandas-dev/pandas: Pandas* (v2.2.2) [Computer software]. Zenodo. https://doi.org/10.5281/zenodo.10957263

Thiebaut de Schotten, M., & Forkel, S. J. (2022). The emergent properties of the connected

brain. *Science*, *378*(6619), 505–510. https://doi.org/10.1126/science.abq2591

Tomczak, M., & Tomczak, E. (2014). *The need to report effect size estimates revisited. An overview of some recommended measures of effect size. 1*.

Vieira, S., Liang, X., Guiomar, R., & Mechelli, A. (2022). Can we predict who will benefit from cognitive-behavioural therapy? A systematic review and meta-analysis of machine learning studies. *Clinical Psychology Review*, *97*, 102193. https://doi.org/10.1016/j.cpr.2022.102193

Virolainen, S. J., VonHandorf, A., Viel, K. C. M. F., Weirauch, M. T., & Kottyan, L. C. (2023). Gene–environment interactions and their impact on human health. *Genes & Immunity*, *24*(1), 1–11. https://doi.org/10.1038/s41435-022-00192-6

Virtanen, P., Gommers, R., Oliphant, T. E., Haberland, M., Reddy, T., Cournapeau, D., Burovski, E., Peterson, P., Weckesser, W., Bright, J., van der Walt, S. J., Brett, M., Wilson, J., Millman, K. J., Mayorov, N., Nelson, A. R. J., Jones, E., Kern, R., Larson, E., … van Mulbregt, P. (2020). SciPy 1.0: Fundamental algorithms for scientific computing in Python. *Nature Methods*, *17*(3), 261–272. https://doi.org/10.1038/s41592-019-0686-2

Waskom, M. L. (2021). seaborn: Statistical data visualization. Journal of Open Source Software, 6(60), 3021. https://doi.org/10.21105/joss.03021